\documentclass[aps,pra,preprint,superscriptaddress,showpacs]{revtex4}

\usepackage{graphicx}
\usepackage{bm}        
\usepackage{amsmath}
\usepackage{amssymb}   

\def\jcp#1#2#3{J.~Chem.~Phys.~{\bf #1},\ #2\ (#3)}
\def\cpl#1#2#3{Chem.~Phys.~Lett.~{\bf #1},\ #2\ (#3)}

\def\pra#1#2#3{Phys.~Rev.~A~{\bf #1},\ #2\ (#3)}
\def\prl#1#2#3{Phys.~Rev.~Lett.~{\bf #1},\ #2\ (#3)}

\def\rmp#1#2#3{Rev.~Mod.~Phys.~{\bf #1},\ #2\ (#3)}

\newcommand{\half}{\frac{1}{2}}
\newcommand{\threejm}[6]{ \left(\begin{array}{ccc} #1 & #3 & #5\\
                                              #2 & #4 & #6
                                \end{array}
                          \right)}

\newcommand{\R}{{\bm R}}
\newcommand{\rA}{\hat{r}_\text{A}}
\newcommand{\rB}{\hat{r}_\text{B}}

\newcommand{\iA}{\tau_\text{A}}
\newcommand{\iB}{\tau_\text{B}}

\newcommand{\gA}{\gamma_\text{A}}
\newcommand{\gB}{\gamma_\text{B}}
\newcommand{\bra}{\langle}
\newcommand{\ket}{\rangle}
\def\lA{\lambda_\text{A}}
\def\lB{\lambda_\text{B}}

\newcommand{\hR}{\hat{R}}

\newcommand{\hrA}{\hat{r}_\text{A}}
\newcommand{\hrB}{\hat{r}_\text{B}}
\renewcommand{\i}{\text{i}}

\begin{document}

\title{Magnetic field modification of ultracold molecule-molecule collisions}

\author{T. V. Tscherbul}
\affiliation{Harvard-MIT Center for Ultracold Atoms, Cambridge, Massachusetts 02138, USA}
\affiliation{Institute for Theoretical Atomic, Molecular, and Optical Physics,
Harvard-Smithsonian Center for Astrophysics, Cambridge, Massachusetts 02138, USA}\email[]{tshcherb@cfa.harvard.edu}
\author{Yu. V. Suleimanov}
\affiliation{Department of Chemistry, Moscow State University, Moscow, 119991, Russia}
\author{V. Aquilanti}
\affiliation{Dipartimento di Chimica, Universit{\`a} di Perugia, Perugia, 06123, Italy}
\author{R. V. Krems}
\affiliation{Department of Chemistry, University of British Columbia, Vancouver, V6T 1Z1, Canada}

\date{\today}

\begin{abstract}
We present an accurate quantum mechanical study of molecule-molecule collisions in the presence of a magnetic field.
The work focusses on the analysis of elastic scattering and spin relaxation in collisions of O$_2(^3\Sigma^-_g)$ molecules
at cold ($\sim$0.1 K) and ultracold ($\sim$10$^{-6}$ K) temperatures. Our calculations show that magnetic spin relaxation in molecule-molecule collisions
is extremely efficient except at magnetic fields below 1 mT. The rate constant for spin relaxation at $T=0.1$ K and a magnetic field of 0.1 T
is found to be as large as $6.1 \times 10^{-11}$ cm$^3$/s. The magnetic field dependence of elastic and inelastic scattering cross sections at ultracold temperatures is dominated by a manifold of Feshbach resonances with the density of $\sim$100 resonances per Tesla for collisions of molecules
in the absolute ground state. This suggests that the scattering length of ultracold molecules in the absolute ground state can be effectively
tuned in a very wide range of magnetic fields. Our calculations demonstrate that the number and properties of the magnetic Feshbach resonances are dramatically different for molecules in the absolute ground and excited spin states. The density of Feshbach resonances for molecule-molecule
scattering in the low-field-seeking Zeeman state is reduced by a factor of 10.

\end{abstract}


\maketitle

\section{Introduction}

The experimental work with cold and ultracold molecules is predicted to lead to many 
fundamental applications in quantum computation \cite{Rabl},
condensed-matter physics \cite{Demler}, precision spectroscopy \cite{Precision}, and physical chemistry \cite{RomanReview}. Recent theoretical work has  shown that ultracold ensembles of molecules trapped in optical lattices can be used to develop novel schemes for quantum information processing \cite{DeMille}, design experiments for quantum simulations of condensed matter physics \cite{Demler,Bloch}, engineer novel phases with topological order \cite{Zoller}, and explore many-body dynamics of strongly interacting systems \cite{Efimov}. Polar molecules in external field traps may form chains, which can be used to study rheological phenomena with non-classical behavior \cite{Demler}.  The creation of a Bose-Einstein condensate of molecules may enable the study of Bose-enhanced chemistry \cite{ami} and the effects of symmetry breaking on chemical interactions at ultracold temperatures \cite{ourPRL}.  The realization of these proposals as well as the creation of dense ensembles of ultracold molecules depends critically on
the possibility of controlling binary molecule - molecule interactions in molecular gases by external electromagnetic fields.
The discovery of magnetic Feshbach resonances in atomic collisions \cite{Verhaar,Ketterle}
opened the door to many groundbreaking experiments with ultracold gases such as
the realization of the BEC-BCS crossover \cite{Jin}, the observation of quantum phase transitions \cite{Bloch} and the creation of ultracold molecules using time-varying magnetic fields \cite{JulienneFeshbach}. Extension of this work to molecular collisions may similarly lead to the development of new research directions such as cold controlled chemistry
\cite{RomanReview}, quantum coherent control of bimolecular reactions \cite{Moshe}, and quantum condensed-matter physics with molecular condensates
\cite{DDI}.

Fueled by the promise of new discoveries, the experimental research of cold and ultracold molecular collisions is expanding rapidly.
Some recent landmark experiments include the observation of threshold collision 
laws in Xe-OH scattering using slow molecular beams \cite{Meijer}, measurements of cross sections for 
D$_2$-OH collisions in a magnetic trap \cite{Ye}, and the detection of magnetic Feshbach resonances in collisions
of Cs$_2$ molecules \cite{Chin}. Further progress in the experimental study of  complex molecule-molecule collisions and chemical reactions at cold and ultracold temperatures requires rigorous theoretical calculations elucidating the mechanisms of energy transfer in ultracold molecular collisions.
Theoretical studies of molecule-molecule collisions at low temperatures are also necessary to understand the prospects for evaporative cooling of molecules in a magnetic trap. The evaporative cooling 
relies on elastic molecule - molecule collisions as the trap depth is gradually reduced
\cite{Ketterle}. In order to sustain efficient evaporative
cooling down to quantum degeneracy, the ratio of the probabilities for elastic scattering and spin relaxation ($\gamma$) in molecule - molecule collisions must exceed $10^4$ \cite{Ketterle}. 
The magnitudes of the rate constants for elastic and inelastic collisions and the dependence of the scattering observables in molecule - molecule encounters on the magnetic field are, however, completely unknown.

Previous theoretical work by Krems and Dalgarno \cite{Roman2004} and Volpi and Bohn \cite{VolpiBohn} identified the main mechanisms of spin relaxation in atom-molecule collisions. It was found that spin relaxation in collisions of $^3\Sigma$ molecules with He atoms
occurs through coupling to rotationally excited states mediated by the spin-spin interaction \cite{Roman2004}.
While the mechanisms of spin relaxation in molecule-molecule collisions should be similar, an accurate computational study
of molecule-molecule scattering in a magnetic field is urgently needed to elucidate the rates of energy relaxation in dense ensembles of trapped
molecules and understand the prospects for evaporative cooling of molecules.
The quantum dynamics of ultracold molecule-molecule collisions in the absence of external fields has been studied by several authors
(see, {\it e.g.}, Refs. \cite{Takayanagi,Alexander,HF}). Avdeenkov and Bohn calculated the rates of spin relaxation in collisions of O$_2(^3\Sigma_g^-)$
molecules at zero magnetic field and observed Feshbach resonances due to molecular rotation \cite{AvdeenkovBohn,RotFR}. 
Krems and Dalgarno \cite{Roman2004} presented a formalism for quantum scattering calculations of cross sections for
molecule-molecule collisions in a magnetic field. However,
they did not consider the symmetrization procedure required to properly describe collisions between identical molecules.
Here, we extend the approach of Krems and Dalgarno to study collisions of identical $^3\Sigma$ molecules.
The use of exchange symmetry reduces the number of scattering channels by a factor of two, which allowed us to obtain
converged cross sections for elastic scattering and spin relaxation of O$_2(^3\Sigma^-_g)$ molecules in an external magnetic field.

We consider ultracold collisions of O$_2$ molecules in their ground electronic state of $^3\Sigma_g^-$ symmetry.
O$_2$--O$_2$ collisions play an important role in atmospheric chemistry \cite{rovib1,rovib2}. The O$_2$--O$_2$ dimer
is an interesting molecular complex \cite{VanDerAvoird1,VanDerAvoird2,JACS,prlO2,rovib1,rovib2,Lamoneda,review,Bartolomei}
(for a recent review, see Ref. \cite{review}). The binding in the complex is affected
by the Heisenberg exchange interaction and the formation of incipient chemical bond \cite{JACS,prlO2}.
Molecular beam scattering experiments \cite{JACS,prlO2} and high-level {\it ab initio} calculations \cite{Lamoneda}
have previously been used to elucidate the nature of the chemical bond in the O$_2$--O$_2$ dimer.
Here, we calculate the magnetic field dependence of the probabilities for spin relaxation in O$_2$--O$_2$ collisions at both
ultracold (down to $10^{-6}$ K) and cold (0.5 K) temperatures relevant for magnetic trapping experiments.
Our calculations elucidate the possibility of evaporative cooling
of paramagnetic $^3\Sigma$ molecules in a magnetic trap and indicate that spin depolarization in molecular collisions
can be controlled by an external magnetic field. The results of our work will be useful for the interpretation of collision
experiments involving slow beams of O$_2$ molecules produced by Zeeman deceleration \cite{Ed} and cryogenic cooling
\cite{DavePatterson}.


\section{Theory}


\subsection{Hamiltonian and symmetrized basis functions}

The Hamiltonian for two $^3\Sigma$ molecules in an external magnetic field can be written as \cite{Roman2004,AvdeenkovBohn}
\begin{equation}\label{H}
\hat{H} = -\frac{1}{2\mu R}\frac{\partial^2}{\partial R^2}R + \frac{\hat{\ell}^2}{2\mu R^2}
+ \hat{V}_\text{el}(\R, {\bm r}_\text{A} ,{\bm r}_\text{B}) 
+ \hat{H}_\text{as},
\end{equation}
where the vectors ${\bm r}_\text{A}$ and ${\bm r}_\text{B}$ describe the orientation of the molecules A and B in the space-fixed
frame, ${\bm R}$ is the vector joining the molecules, $\hat{\ell}$ is the orbital angular momentum of the collision complex, $\mu$
is the reduced mass, and $\hat{V}_\text{el}$ is the electrostatic interaction potential (see Sec. IIB). 
The asymptotic Hamiltonian $\hat{H}_\text{as}$
describes the individual molecules A and B in the presence of a magnetic field \cite{Roman2004}. 
\begin{equation}\label{Has}
\hat{H}_\text{as} = \hat{H}_\text{A} + \hat{H}_\text{B},
\end{equation}
The Hamiltonians of an individual $^3\Sigma^-$ molecule can be written as \cite{Mizushima}
\begin{equation}\label{Has2}
 \hat{H}_\nu = B_{e}\hat{N}_\nu^2 +2\mu_\text{B}\bm{B}\cdot \hat{S}_\nu
+ \gamma \hat{N}_\nu \cdot \hat{S}_\nu
  + \frac{2}{3}\lambda_\text{SS}\left( \frac{24\pi}{5}\right)^{1/2}\sum_{q} Y_{2q}^\star (\hat{r}_\nu) [\hat{S}_\nu\otimes \hat{S}_\nu]^{(2)}_q,
\end{equation}
where the index $\nu=\text{A},\text{B}$ labels the molecule, $\hat{N}_\nu^2$ is the rotational angular momentum, $B_e$ is the rotational constant,
$\mu_\text{B}$ is the Bohr magneton, ${\bm B}$ is the magnetic field vector and
 $\hat{S}_\nu$ is the electron spin. The spin-rotation and spin-spin interactions in Eq. (\ref{Has2}) are parametrized by
the phenomenological constants $\gamma$ and $\lambda_\text{SS}$, which do not depend on $\nu$ since we
consider collisions of identical molecules. The spin-spin interaction given by the last term in Eq. (\ref{Has2})
can be represented as a sum of tensor products of two identical spin operators and
spherical harmonics, which depend on the orientation $\hat{r}_\nu$ of the diatomic molecule in the space-fixed (SF) coordinate frame \cite{Mizushima}.
We use the rigid rotor approximation $({\bm r}_\nu =  \hat{r}_\nu)$ and neglect the weak magnetic dipole-dipole interaction and
the hyperfine interaction of~$^{17}$O \cite{AvdeenkovBohn}.

We solve the scattering problem for two identical bosonic molecules by expanding the total wave function in a space-fixed uncoupled basis
introduced in Ref. \cite{Roman2004}. The uncoupled basis functions are products of the eigenfunctions of the operators $\hat{N}_\nu^2$ and $\hat{S}_\nu^2$,
their projections along the space-fixed $z$-axis $\hat{N}_{\nu_z}$ and $\hat{S}_{\nu_z}$, and the spherical harmonics $|\ell m_\ell\rangle$
\begin{equation}\label{UnsymmetrizedBasis}
| \iA \iB  \ell m_\ell \rangle = | \iA \rangle |\iB \rangle | \ell m_\ell \rangle
\end{equation}
where $|\tau_\nu\rangle = |N_\nu M_{N_\nu}\rangle |S_\nu M_{S_\nu}\rangle$. The matrix elements of the Hamiltonians (\ref{H}), (\ref{Has}) in the basis (\ref{UnsymmetrizedBasis}) were derived by Krems and Dalgarno \cite{Roman2004}
for the case of distinguishable molecules. The basis functions for identical molecules must be the eigenfunctions of the permutation operator
$\hat{P}: {\bm r}_\text{A} \to {\bm r}_\text{B}; {\bm r}_\text{B} \to {\bm r}_\text{A}; \R \to -{\bm R}$, which commutes with the total Hamiltonian (\ref{H}).
The symmetrized basis functions can be constructed by applying the operator $1+\hat{P}$ to Eq.~(\ref{UnsymmetrizedBasis}) and normalizing
the result \cite{Roman2004,Green,Kouri, Alexander}
\begin{equation}\label{SymmetrizedBasis}
\phi_{\iA \iB \ell m_\ell}^{\eta\epsilon}
= \frac{1}{[2(1+\delta_{\iA \iB})]^{1/2}} 
\bigl{[} |\iA \iB\rangle + \eta \epsilon |\iB \iA\rangle \bigr{]} |\ell m_\ell \rangle
\end{equation}
where $\epsilon = (-)^\ell$ and the symmetry of the basis functions with respect to the interchange of identical molecules is given
by index $\eta$ which is equal to 1 for composite bosons and $-1$ for composite fermions~\cite{SoldanHutsonIRPC}.
In what follows we assume that the first basis function in each product on the right-rand side of Eq. (\ref{SymmetrizedBasis}) depends
on the coordinates of molecule A,
and the second basis function depends on the coordinates of molecule B \cite{footnote1}.
Equaton~(\ref{SymmetrizedBasis}) defines the well-ordered (combined) molecular states with
$\tau_\text{A}\ge \tau_\text{B}$.
The normalization factor $[2(1+\delta_{\iA\iB})]^{1/2}$ takes into account
the fact that the basis function (\ref{SymmetrizedBasis}) with $\iA=\iB$ need not be symmetrized when
$\eta\epsilon=+1$ (i.e. when the colliding molecules are in the same state). The symmetrized
function (\ref{SymmetrizedBasis}) with $\iA=\iB$ vanishes identically when $\eta\epsilon = -1$.

The basis functions given by Eqs. (\ref{UnsymmetrizedBasis}) and (\ref{SymmetrizedBasis}) are also the
eigenfunctions of the inversion operator, with eigenvalues given by $(-)^{N_\text{A}+N_\text{B}+\ell}$ \cite{Kouri}.
Since for homonuclear molecules both $N_\text{A}$ and $N_\text{B}$ are either even or odd,
the quantity $\epsilon=(-)^\ell$ is conserved as well. Thus, it is convenient to rewrite Eq. (\ref{SymmetrizedBasis}) in
a factorized form  
\begin{equation}\label{I1}
\phi_{\iA \iB \ell m_\ell}^{\eta\epsilon} =  \mathcal{I}_{\iA\iB}^{\eta\epsilon}(\rA,\rB)
|\ell m_\ell\rangle
\end{equation}
where
\begin{equation}\label{I2}
\mathcal{I}_{\iA\iB}^{\eta\epsilon}(\rA,\rB) = \frac{1}{[2(1+\delta_{\iA \iB})]^{1/2}}
\bigl{[} |\iA \iB\rangle + \eta\epsilon |\iB \iA\rangle \bigr{]}.
\end{equation}
We note that no such symmetry exists for heteronuclear molecules, where the basis functions with $\Delta \ell = \pm 1$
are coupled by the interaction potential, and $\epsilon$ is not conserved.


The uncoupled symmetrized basis functions given by Eq. (\ref{SymmetrizedBasis}) are not the eigenfunctions
of the asymptotic Hamiltonian (\ref{Has}) because the spin-rotation and spin-spin interactions couple the states
with different $N$, $M_N$, and $M_S$. Since the scattering $S$-matrix must be defined in terms of the eigenfunctions
of $\hat{H}_\text{as}$, it is necessary to transform the wave function to a new basis \cite{Roman2004} 
\begin{equation}\label{Transformation1}
\mathcal{I}_{\gA \gB}^{\eta\epsilon}(\rA,\rB) =
\sum_{\iA\ge\iB} C_{\iA\iB, \gA\gB}^{\eta\epsilon} \mathcal{I}_{\iA \iB}^{\eta\epsilon}(\rA,\rB),
\end{equation}
where the coefficients $C_{\iA\iB, \gA\gB}^{\epsilon\eta}$ form the matrix, which diagonalizes the asymptotic Hamiltonian
in the symmetrized basis~(\ref{SymmetrizedBasis}). More explicitly, $\mathsf{C}^\text{T}\mathsf{H}_\text{as}\mathsf{C} = \mathsf{E}$,
where $\mathsf{E}$ is the diagonal matrix of asymptotic energies and $\epsilon_{\gA\gB} = \epsilon_{\gA}+\epsilon_{\gB}$.
The energies of individual molecules $\epsilon_\nu$ can be obtained by diagonalization
of individual molecule Hamiltonians $\hat{H}_\nu$ (\ref{Has}) in the subspace of functions $|\tau_\nu\rangle$.
Exploiting the properties of the interchange operator $\hat{P}^\dagger = \hat{P}$ and $\hat{P}^2=1$ and using
Eq.~(\ref{SymmetrizedBasis}), we obtain the following expression for the matrix elements of $\hat{H}_\text{as}$
\begin{multline}
\bra \phi_{\iA\iB\ell m_\ell}^{\eta\epsilon}|\hat{H}_\text{as}| \phi_{\iA'\iB'\ell' m_\ell'}^{\eta\epsilon}\ket
= \frac{\delta_{\ell\ell'}\delta_{m_\ell m_\ell'}}{ [{(1+\delta_{\iA\iB})(1+\delta_{\iA'\iB'})}]^{1/2} }
\bigr{[}\delta_{\iB,\iB'}\bra \iA| \hat{H}_\text{A}|\iA'\rangle 
+\eta \epsilon \delta_{\iB,\iA'}\bra \iA| \hat{H}_\text{A}|\iB'\rangle 
\bigr{]}.
\end{multline}
This expression provides a convenient method of constructing the matrix elements of the asymptotic Hamiltonian (\ref{Has})
in the symmetrized basis. Alternatively, one can obtain the transformation
coefficients (\ref{Transformation1}) directly from the eigenvectors of $\hat{H}_\text{A}$ defined as
\begin{equation}\label{gA}
|\gA \ket=\sum_{\iA}C_{\iA \gA}|\iA\ket.
\end{equation}
The transformation can be derived by multiplying Eq. (\ref{gA}) by a similar expression for $|\gB \ket$ and rearranging the terms.
The result is
\begin{equation}\label{Connection}
C_{\iA\iB, \gA\gB}^{\eta\epsilon} = 
\frac{1}{[(1+\delta_{\iA,\iB})(1+\delta_{\gA,\gB})]^{1/2}}
[C_{\iA \gA}C_{\iB \gB}
+ \eta \epsilon C_{\iA \gB}C_{\iB \gA}]
\end{equation}
We have verified by numerical tests that Eqs. (\ref{Transformation1}) and (\ref{Connection}) give identical
results up to an unimportant overall phase. In practice, however, it is more convenient to use Eq. (\ref{Connection}) because it
requires the evaluation and the diagonalization of smaller matrices.

\subsection{Interaction Potential}

The spin-dependent interaction potential between two $^3\Sigma$ molecules can be represented in the form
\begin{equation}\label{VS}
\hat{V}_\text{el}({\bm R},\rA,\rB) = \sum_{S=-1}^1\sum_{M_S=-S}^S V_S({\bm R},\rA,\rB) |SM_S\rangle \langle SM_S|,
\end{equation}
where $\hat{S} = \hat{S}_\text{A}+\hat{S}_\text{B}$ is the total spin and $M_S=M_{S_\text{A}} + M_{S_\text{B}}$.
The representation (\ref{VS}) is often used to describe the interactions of ultracold alkali metal atoms \cite{Verhaar}.
Here, we choose an alternative approach proposed by van der Avoird and coworkers \cite{VanDerAvoird1,VanDerAvoird2},
in which the full interaction potential is separated into the spin-independent and spin-dependent parts 
\begin{equation}\label{Vseparation}
\hat{V}_\text{el}({\bm R},\rA,\rB) = \hat{V}_\text{si}({\bm R},\rA,\rB) + \hat{V}_\text{sd}({\bm R},\rA,\rB).
\end{equation}
The spin-dependent part, also known as the Heisenberg exchange interaction \cite{VanDerAvoird1,VanDerAvoird2}, can be parametrized
for two $^3\Sigma$ molecules as
\begin{equation}\label{Heisenberg}
\hat{V}_\text{sd}({\bm R},\rA,\rB) = -2J({\bm R},\rA,\rB)\hat{S}_\text{A}\cdot\hat{S}_\text{B},
\end{equation}
where $J({\bm R},\rA,\rB)$ is a scalar function similar to the
interaction potential but decreasing exponentially with $R$ \cite{VanDerAvoird2}. We expand
the interaction potential in the angular basis \cite{Roman2004,VanDerAvoird1,footnote2}
\begin{equation}\label{V}
V_\text{si}({\bm R},\rA,\rB)
= (4\pi)^{3/2}\sum_{\lambda_\text{A},\lambda_\text{B},\lambda}V_{\lA\lB\lambda}(R) A_{\lA\lB\lambda}(\hR,\hrA,\hrB),
\end{equation}
with the space-fixed angular basis functions defined as
\begin{equation}\label{A}
A_{\lA\lB\lambda}(\hR,\hrA,\hrB)=
\sum_{m_{\lA},m_{\lB},m_\lambda}\threejm{\lA}{m_{\lA}}{\lB}{m_{\lB}}{\lambda}{m_{\lambda}}Y_{\lA m_{\lA}}(\hrA)
Y_{\lB m_{\lB}}(\hrB)Y_{\lambda m_\lambda}(\hR),
\end{equation}
where (:::) denotes a 3-$j$ symbol. An efficient procedure for the evaluation of the radial expansion coefficients in
Eq. (\ref{V}) is described in Appendix A.


The matrix elements of the spin-indepenent interaction potential between the symmetrized
basis functions (\ref{SymmetrizedBasis}) have the form
\begin{align}\label{MatrixElementV1}
\langle \phi_{\iA\iB\ell m_\ell}^{\eta\epsilon} |V_\text{si}| \phi_{\iA'\iB'\ell'm_\ell'}^{\eta\epsilon}\rangle
&= \frac{1}{[(1+\delta_{\iA\iB})(1+\delta_{\iA'\iB'})]^{1/2}} \\ &\times \notag
\bigl{[}   \langle {\iA} {\iB} \ell m_\ell|  V_\text{si}  | {\iA'} {\iB'}\ell'  m_\ell' \rangle 
+ \eta\epsilon\langle \iA \iB \ell m_\ell | V_\text{si} | \iB' \iA' \ell' m_\ell'\rangle \bigl{]}
\end{align}
The first (direct) term on the right-hand side is a matrix element in the unsymmetrized basis.
Krems and Dalgarno \cite{Roman2004} evaluated this matrix element using the
expansion of $V_\text{si}$ in terms of the angular basis functions (\ref{V}). Using the conservation of the total angular momentum
projection $M=M_{N_\text{A}}+M_{S_\text{A}}+M_{N_\text{B}}+M_{S_\text{B}}+m_\ell$ for collisions in a magnetic
field, we obtain
\begin{multline}\label{DirectMatrixElement}
\langle {\iA} {\iB} \ell m_\ell | V_\text{si} | {\iA'} {\iB'} \ell' m_\ell'\rangle
=\delta_{M_{S_\text{A}},M_{S_\text{A}'}} \delta_{M_{S_\text{B}},M_{S_\text{B}'}}\sum_{\lA,\lB,\lambda}
V_{\lA\lB\Lambda}(R) [(2N_\text{A}+1)(2N_\text{A}'+1) \\ \times 
(2N_\text{B}+1)(2N_\text{B}'+1)(2\ell+1)(2\ell'+1)
(2\lA + 1)(2\lB+1)(2\lambda+1)]^{1/2}(-)^{ M_{N_\text{A}} + M_{N_\text{B}} + m_\ell } \\ \times
\threejm{\lA}{ M_{N_\text{A}} - M_{N_\text{A}}' }{\lB} { M_{N_\text{B}} - M_{N_\text{B}}' }{\lambda}{ m_\ell - m_\ell' }  
\threejm{N_\text{A}} {-M_{N_\text{A}}} {\lA}{M_{N_\text{A}}-M_{N_\text{A}}'} {N_\text{A}'} {M_{N_\text{A}}'}
\threejm{N_\text{A}} {0} {\lA}{0} {N_\text{A}'} {0}  \\ \times
\threejm{N_\text{B}} {-M_{N_\text{B}}} {\lB}{M_{N_\text{B}}-M_{N_\text{B}}'} {N_\text{B}'} {M_{N_\text{B}}'}
\threejm{N_\text{B}} {0} {\lB}{0} {N_\text{B}'} {0}
\threejm{\ell} {-m_\ell} {\lambda}{m_\ell - m_\ell'} {\ell'} {m_\ell'} 
\threejm{\ell} {0} {\lambda}{0} {\ell'} {0}
\end{multline}
The second (exchange) matrix element appears as a result of the symmetrization procedure
(\ref{SymmetrizedBasis}). It can be obtained from Eq. (\ref{DirectMatrixElement}) by interchanging
the indexes $N_\text{A}'\leftrightarrow N_\text{B}'$, $M_{N_\text{A}}' \leftrightarrow M_{N_\text{B}}'$, and $M_{S_\text{A}}' \leftrightarrow M_{S_\text{B}}'$.

The matrix elements of the Heisenberg exchange interaction are obtained
using the same procedure:
\begin{align}\label{MatrixElementJ}
\langle \phi_{\iA\iB\ell m_\ell}^{\eta\epsilon} |\hat{V}_\text{ex}&| \phi_{\iA'\iB'\ell' m_\ell'}^{\eta\epsilon}\rangle
= \frac{1}{[(1+\delta_{\iA\iB})(1+\delta_{\iA'\iB'})]^{1/2}} \\ &\times \notag
\bigl{[} \langle S_\text{A} M_{S_\text{A}} | \langle S_\text{B} M_{S_\text{B}} | \hat{S}_\text{A}\cdot \hat{S}_\text{B} 
|S_\text{A} M_{S_\text{A}}'\rangle |S_\text{B} M_{S_\text{B}}'\rangle
\langle {\iA} {\iB} \ell m_\ell | -2J_\text{AB} | {\iA'} {\iB'} \ell' m_\ell'\rangle \\ &+ \notag
\eta \epsilon
\langle S_\text{A} M_{S_\text{A}} | \langle S_\text{B} M_{S_\text{B}} | \hat{S}_\text{A}\cdot\hat{S}_\text{B} 
|S_\text{B} M_{S_\text{B}}'\rangle |S_\text{A} M_{S_\text{A}}'\rangle
\langle \iA \iB \ell m_\ell | -2J_\text{AB} | \iB' \iA' \ell' m_\ell'\rangle \bigl{]}.
\end{align}
The matrix elements of the prefactor $-2J(R,\theta_\text{A},\theta_\text{B},\varphi)$ are given by
Eq. (\ref{DirectMatrixElement}) with $V_{\lA\lB\lambda}(R)$ replaced by $-2J_{\lA\lB\lambda}(R)$.
Equation (\ref{MatrixElementJ}) shows that unlike the spin-independent matrix elements (\ref{MatrixElementV1}),
both the direct and exchange contributions to the Heisenberg exchange matrix element are not diagonal
in spin quantum numbers $M_{S_\text{A}}$ and $M_{S_\text{B}}$. For completeness, we give
here the expressions for the matrix elements of the operator
$\hat{S}_\text{A}\cdot\hat{S}_\text{B}$.
The direct matrix element has the form \cite{Roman2004}
\begin{multline}\label{SASB}
\langle S_\text{A} M_{S_\text{A}} | \langle S_\text{B} M_{S_\text{B}} | \hat{S}_\text{A}\cdot\hat{S}_\text{B} 
|S_\text{A} M_{S_\text{A}}'\rangle |S_\text{B} M_{S_\text{B}}'\rangle = \delta_{M_{S_\text{A}} M_{S_\text{A}}'} 
\delta_{M_{S_\text{B}} M_{S_\text{B}}'}M_{S_\text{A}} M_{S_\text{B}} \\ +
\textstyle \half \delta_{M_{S_\text{A}}, M_{S_\text{A}}'\pm 1} \delta_{M_{S_\text{B}}, M_{S_\text{B}}'\mp 1} 
\bigl{[}S_\text{A}(S_\text{A}+1) - M_{S_\text{A}}'(M_{S_\text{A}}' \pm 1)\bigr{]}^{1/2}
\bigl{[}S_\text{B}(S_\text{B}+1) - M_{S_\text{B}}'(M_{S_\text{B}}' \mp 1)\bigr{]}^{1/2},
\end{multline}
and the exhange matrix element can be obtained from Eq. (\ref{SASB}) 
by interchanging the indexes $M_{S_\text{A}}'$ and $M_{S_\text{B}}'$.

\subsection{Close-coupling equations, scattering amplitudes, and cross sections}

The symmetrized wave function of the collision complex can be expanded as 
\begin{equation}\label{Expansion2}
\psi^\eta =\frac{1}{R} \sum_{\gA\ge \gB} \sum_{\ell,m_\ell}
F^\eta_{\gA\gB\ell m_\ell}(R) Y_{\ell m_\ell}(\hR) \mathcal{I}^{\eta\epsilon}_{\gA\gB}(\rA,\rB).
\end{equation} 
Substituting this expansion into the Schr\"odinger equation,
$H\psi^\eta = E\psi^\eta$, where $E$ is the total energy,
we obtain the close-coupling equations for the radial expansion coefficients
\begin{equation}\label{CC}
\Bigl{[} \frac{d^2}{dR^2} - \frac{\ell(\ell+1)}{R^2}+2\mu E\Bigr{]} F^\eta_{\gA\gB\ell m_\ell}(R)
= \sum_{\gA'\gB'}\sum_{\ell'm_\ell'}
\bigl{[}\mathsf{C}^\text{T}\mathsf{V}\mathsf{C}\bigr{]}_{\gA\gB\ell m_\ell;\gA'\gB'\ell'm_\ell'}
 F^\eta_{\gA'\gB'\ell' m_\ell'}(R).
\end{equation}
The asymptotic form of the solutions $F^\eta_{\gA\gB\ell m_\ell;\gA'\gB'\ell'm_\ell'}(R)$ 
at $R\to\infty$ defines the symmetrized scattering matrix
\begin{align}\notag
F^\eta_{\gA\gB\ell m_\ell;\gA'\gB'\ell'm_\ell'}(R\to\infty) &\to \delta_{\gA\gA'}\delta_{\gB\gB'}
\delta_{\ell\ell'}\delta_{m_\ell m_\ell'} \exp[-\i (k_{\gA\gB}R-\pi \ell/2)] \\  \label{DefinitionS}
 &- \biggl{(}\frac{k_{\gA\gB}}{k_{\gA'\gB'}}\biggr{)}^{1/2}S^\eta_{\gA\gB\ell m_\ell;\gA'\gB'\ell'm_\ell'}
\exp[\i(k_{\gA'\gB'}R-\pi \ell'/2)]
\end{align}
The scattering amplitude for indistinguishable molecules can be written as \cite{HF,MottMassey}
\begin{equation}\label{qsymm}
\tilde{q}^\eta_{\gA\gB\to \gA'\gB'}(\hR_\i,\hR) =  q_{\gA\gB \to \gA'\gB'}(\hR_\i,\hR) + \eta q_{\gA\gB \to \gB'\gA'}(\hR_\i,-\hR).
\end{equation}
where the scattering amplitudes on the right-hand side can be written in terms of unsymmetrized $T$-matrix elements
\begin{equation}\label{qT}
q_{\gA\gB \to \gA'\gB'}(\hR_\i,\hR) = 2\pi\sum_{\ell,m_\ell}\sum_{\ell',m_\ell'} \i^{\ell-\ell'}
Y^*_{\ell m_\ell}(\hR_\i) 
Y_{\ell' m_\ell'}(\hR) T_{\gA\gB \ell m_\ell ; \gA'\gB' \ell' m_\ell'}.
\end{equation}
Substituting this expression into Eq. (\ref{qsymm}) and rearranging the terms, we find
\begin{multline}\label{qT}
\tilde{q}^\eta_{\gA\gB \to \gA'\gB'}(\hR_\i,\hR)=
2\pi[(1+\delta_{\gA\gB})(1+\delta_{\gA'\gB'})]^{1/2} 
\sum_{\ell,m_\ell}\sum_{\ell',m_\ell'} \i^{\ell-\ell'} 
Y^*_{\ell m_\ell}(\hR_\i) 
Y_{\ell' m_\ell'}(\hR)  \\ \times
T^\eta_{\gA\gB \ell m_\ell ; \gA'\gB' \ell' m_\ell'},
\end{multline}
where 
\begin{equation}\label{T}
T^\eta_{\gA\gB \ell m_\ell ; \gA'\gB' \ell' m_\ell'}
=\frac{1}{[(1+\delta_{\gA\gB})(1+\delta_{\gA'\gB'})]^{1/2}}
\bigl{[}
T_{\gA\gB \ell m_\ell ; \gA'\gB' \ell' m_\ell'} + \eta\epsilon T_{\gA\gB \ell m_\ell ; \gB'\gA' \ell' m_\ell'}
\bigr{]}.
\end{equation}
These $T$-matrix elements are exactly the same as those obtained from the solution of the close-coupling equations
in the symmetrized basis set (as described above). They are related to the $S$-matrix elements
via \cite{Dalgarno}
\begin{equation}\label{S2T}
T^\eta_{\gA\gB \ell m_\ell ; \gA'\gB' \ell' m_\ell'}
= \delta_{\gA\gA'}\delta_{\gB\gB'}\delta_{\ell\ell'}\delta_{m_\ell m_\ell'}
-S^\eta_{\gA\gB \ell m_\ell ; \gA'\gB' \ell' m_\ell'}.
\end{equation}

The integral cross section for indistinguishable molecules in fully polarized nuclear spin states can be obtained
from the scattering amplitude as described in Appendix B. The final result is
\begin{equation}\label{ICSfinal}
\sigma_{\gA\gB\to\gA'\gB'} =
\frac{\pi (1+\delta_{\gA\gB})}
{k^2_{\gA\gB}} \sum_{\ell,m_\ell}\sum_{\ell' m_\ell'}|T^\eta_{\gA\gB\ell m_\ell ; \gA'\gB'\ell' m_\ell'}|^2,
\end{equation}
where $k^2_{\gA\gB} = 2\mu(E-\epsilon_{\gA} - \epsilon_{\gB})$ is the wave vector for the initial collision channel.
This equation shows that the elastic cross section
of indistinguishable bosons (or fermions) is twice as large as that of distinguishable particles.
Our Eq. (\ref{ICSfinal}) agrees with the results presented by Avdeenkov and Bohn \cite{AvdeenkovBohn} and
Burke \cite{Burke}.



\subsection{Computational details}

To parametrize the asymptotic Hamiltonian (\ref{Has}), we used the accurate spectroscopic constants of $^{17}$O$_2$ (in units of cm$^{-1}$):
$B_e = 1.353$, $\gamma = -0.00396$, $\lambda_\text{SS} = 1.985$ \cite{Gazzoli}.
In order to obtain converged cross sections at collision energies below 0.5 K,
we used a basis set comprising 3 rotational states ($N_\text{A},\, N_\text{B}=0-4$) and 4 partial waves ($\ell=0-6$). These parameters resulted
in 2526 coupled differential equations for $M=0$. At the lowest collision energy of $10^{-6}$ K, four partial
waves were sufficient to achieve convergence. The coupled equations (\ref{CC}) were solved on a grid of $R$ from 2~$a_0$ to 200 $a_0$
with a step size of $0.04$ $a_0$ yielding the cross sections converged to within 10\%.
The spin-independent and spin-dependent parts of the interaction potential for the oxygen dimer were constructed
to reproduce the integral cross sections measured in molecular beam scattering experiments \cite{JACS,prlO2}.
The lowest-order isotropic part $V_{000}(R)$ was determined with high accuracy from an analysis of the glory pattern 
in high-velocity collisions of rotationally ``hot'' O$_2$ molecules. The anisotropic coefficients $V_{202}(R)$,
$V_{220}(R)$, and $V_{222}(R)$ were inferred from the scattering experiments with rotationally cold, supersonic beams of aligned
O$_2$ \cite{JACS,prlO2}. Recent {\it ab initio} studies \cite{Bartolomei,PCCP} have shown that the O$_2$--O$_2$ interaction
may be more anisotropic than suggested in Ref. \cite{JACS}. The inelastic transition probabilities in molecule-molecule collisions increase with
the anisotropy of the interaction potential \cite{Roman2004}, so the cross sections for spin relaxation presented in this work
should be viewed as lower bounds to the actual magnitudes.



\section{Results and discussion}

We consider collisions of two identical oxygen molecules in the ground electronic state $^3\Sigma_g^-$. We choose the $^{17}$O
isotope because the $^{17}$O$_2$ molecule is characterized by the $J_\text{A}=1$ ground state required for magnetic trapping
(where $\hat{J}_\text{A}=\hat{N}_\text{A}+\hat{S}_\text{A}$ is the total angular momentum of O$_2$ exclusive of nuclear spin).
The Zeeman levels of $^{17}$O$_2$ obtained by diagonalizing the asymptotic Hamiltonian are shown in Fig. 1($a$). Magnetic fields
split the $J_\text{A}=1$ ground state of $^{17}$O$_2$ into three Zeeman states corresponding to $M_{S_\text{A}}=0$, $-1$, and $+1$.
We first consider collisions of two oxygen molecules in the spin-aligned state $|M_{S_\text{A}}=1,M_{S_\text{B}}=1\rangle$,
relevant for magnetic trapping and cryogenic cooling experiments.
Molecules prepared in this state can scatter elastically or undergo spin-changing inelastic transitions. 
The lower panel of Fig. 2 shows that spin-changing collisions are suppressed
at low magnetic fields ($<1$ mT) leading to a large value of the elastic to inelastic ratio $\gamma \sim 10^4$ which is favorable for evaporative cooling.
Since $M$ is conserved, the spin relaxation transitions must be accompanied by the transition
$\ell=0\to \ell'=2$, which leads to a centrifugal barrier in the outgoing reaction channel suppressing the $s$-wave inelastic scattering. Because
the $\ell=2$ barrier has a fixed height of 13 mK \cite{AvdeenkovBohn}, this suppression occurs only at low magnetic fields
where the energy difference between the initial and final Zeeman states does not exceed the barrier height \cite{AvdeenkovBohn,VolpiBohn}.
At larger magnetic fields, the spin-changing transitions are much more efficient, as shown in Fig. 2($b$).
This suggests that evaporative cooling of $^3\Sigma$ molecules in a deep magnetic trap will be challenging.
However, our results indicate that evaporative cooling may be possible in shallow magnetic traps ($<1$ mT deep).
The maximum temperature of the molecules that can be held in such a trap is determined by the parameter 
\begin{equation}
\eta = \frac{|\mu B|}{k_\text{B}T},
\end{equation}
where $T$ is the temperature, $k_\text{B}$ is the Boltzmann's constant, and $\mu = 2\mu_0$ for $^3\Sigma$ molecules.
Efficient trapping requires $\eta>5$ \cite{Doyle}, which corresponds to a temperature of 0.25 mK for $^3\Sigma$ molecules at $B=1$ mT.
Such temperatures can be reached using optical Stark deceleration \cite{Barker} or Zeeman slowing~\cite{Ed}.

Spin relaxation in collisions of $\Sigma$-state molecules
in fully spin-stretched states can only occur through coupling to the rotationally excited states assisted by
the spin-spin interaction \cite{Roman2004}. This is a two-step mechanism illustrated schematically in Fig. 1($b$).
The spin-spin interaction (denoted by $V_\text{SS}$ in Fig. 1) mixes the ground rotational state $|N=0,M_N=0,M_S=1\rangle$
with the excited rotational states $|N=2,M_N,M_S=-1\rangle$. Spin relaxation is then induced by the anisotropy of the interaction potential leading to the
$|N=2,M_N,M_S=-1\rangle \to |N'=0,M_N',M_S'=-1\rangle$ transition. The anisotropy of molecule - molecule interaction potentials is usually very strong; our results yield the rate constant for spin relaxation 
$6.1\times 10^{-11}$ cm$^3$/s at $T=0.1$ K and $B=0.1$ T.

Figure 3 shows the magnetic field dependence of O$_2$ - O$_2$ scattering at a collision
energy of 10$^{-6}$ K. The spin relaxation cross section increases from zero to a large value over a narrow region
of magnetic fields $B=0-1$ mT.
For magnetic fields larger than 10 mT, the spin relaxation cross section summed over all final Zeeman states is
a factor of 10 larger than the elastic scattering cross section. 
The cross section displays several broad resonances. Figure 1($b$) illustrates that
at certain magnetic fields, the energy in the incoming collision channel becomes degenerate with that of
a quasibound state supported by the uppermost curve. This results in the formation of a long-lived complex in which
one of the molecules is in the $N=2$ rotationally excited state \cite{AvdeenkovBohn}.
The lifetime of the complex is determined by the strength of the couplings induced by the spin-rotation interaction and the intermolecular potential as well as
the presence of inelastic loss processes. In particular, the peaks shown in Fig. 3 are relatively
broad because inelastic spin relaxation leads to suppression of the $S$-matrix poles \cite{Jeremy}.

The magnetic field dependence of the scattering cross sections for collisions of  molecules in the absolute ground high-field-seeking state $|M_{S_\text{A}}=-1,M_{S_\text{B}}=-1\rangle$ is dramatically different (see Figure 4).  The $s$-wave elastic scattering cross section
displays a manifold of resonance peaks which we attribute to the combined action of the interaction potential and the spin-spin interaction (\ref{H}).
As a result of an interplay between these interactions, avoided crossings occur between the incoming scattering state
$|M_{S_\text{A}}=-1,M_{S_\text{B}}=-1\rangle$ and the quasibound states of the O$_2$--O$_2$ complex shown in Fig.~1($b$)
leading to the resonant variation of the elastic cross section.
This suggests that the resonances depicted in Fig. 4 are similar to the magnetic Feshbach resonances in collisions of the alkali metal atoms
\cite{Verhaar} and that the $s$-wave scattering length in an ultracold gas of $^3\Sigma$ molecules can be efficiently manipulated with magnetic fields.

In order to elucidate the possibility for evaporative cooling of $^3\Sigma$ molecules, it is useful to analyze the dependence of the
spin relaxation rates on the rotational constant \cite{Roman2004,OurWork}. In Fig. 5, we plot
the ratio of the cross sections for elastic scattering and spin relaxation as a function of the splitting between the
rotational levels. All other parameters of the Hamiltonian were unchanged in this calculation.
The probability for spin relaxation in atom-molecule and molecule-molecule collisions decreases with increasing $B_e$ because
the matrix element between the different spin states due to the spin-spin interaction scales as $\lambda_\text{SS}/B_e$ \cite{Roman2004,OurWork}.
Figure 5 shows that the elastic to inelastic ratio at a collision energy of 0.1 K improves dramatically with
increasing the rotational constant. The dependence on $B_e$ is, however, not monotonic because of the manifold of
Feshbach resonances (see Figs. 3-4), which are most pronounced at low collision energies.
The results shown in Fig. 5 indicate that evaporative cooling of molecules with larger rotational constants should generally be more efficient.

Previous theoretical and experimental work has established that the cross sections for spin-changing transitions in
$^3\Sigma$ molecules are sensitive to the splitting between the $N=0$ and the $N=2$ rotational levels.
The approximate $1/B_e^2$ scaling law \cite{OurWork} suggests that spin relaxation occurs at a slower rate in collisions
of atoms and molecules with larger rotational constants. This result is important because the splitting between
different rotational levels can be altered with electric, microwave, or off-resonant laser fields, thereby enabling control
over the spin degrees of freedom of cold molecules. We illustrate the idea using the example of a nonpolar $^3\Sigma$ molecule
such as O$_2$ in the presence of far off-resonant, continuous-wave laser field. The energy levels are determined by
the Hamiltonian (\ref{Has2}) with an additional term~\cite{Friedrich}
\begin{equation}\label{L}
\hat{V}_\text{L} = -\frac{\epsilon_0^2}{4}\left[ \Delta \alpha \cos^2\chi +\alpha_\perp \right],
\end{equation}
where $\Delta \alpha = \alpha_{||}-\alpha_\perp$ is the polarizability anisotropy, $\chi$ is the angle between the molecular axis and
the laser polarization vector (which we assume to be parallel to ${\bm B}$), and $\epsilon_0$ is the amplitude of the laser field. 
Figure 6 shows the energy levels of $^{17}$O$_2$ as functions of the laser field strength (we transform away the $\chi$-independent terms
by properly choosing the zero of energy). The effective rotational constant defined as the splitting between the closest field-dressed
Zeeman levels in the $N=0$ and $N=2$ manifolds, increases by a factor of 1.6 as the laser field intensity is varied from zero to
$5\times 10^{11}$ W/cm$^2$. For polar molecules, similar effects can be induced by dc and microwave laser fields \cite{OurWork2}.

As shown in Fig. 6, any substantial modification of molecular energy levels requires laser field intensities of order 10$^{11}$ W/cm$^2$.
The maximum cw laser field intensity currently available in the laboratory is about $2.5 \times 10^5$ W/cm$^2$ for a 0.5 mm size sample.
However, if the trapped cloud is compressed to one-tenth its original size, the maximum attainable laser intensity increases
by a factor of 100, so the required field strengths of $10^{11}$ W/cm$^2$ are well within reach for microscopic clouds ($10^{-3}$ mm).
Samples of such size can be produced and stored in magnetic microtraps recently demonstrated for cold polar molecules \cite{MeijerChip}. 
An alternative solution is to use pulsed lasers, whose intensity is typically much stronger than that of cw lasers.
However, the time-independent description used in this work can only be applied if the duration of the laser pulse is long compared to
the collision time. This condition is necessary to ensure the validity of the quasistatic approximation for the molecule-field interaction \cite{Friedrich}.

Figures 7 and 8 show the cross sections for spin relaxation and elastic scattering in O$_2$--O$_2$ collisions as functions of the rotational constant at
different collision energies. At the highest energy of 0.1 K, the cross section for spin relaxation decreases monotonically and that for
elastic scattering increases with increasing $B_e$. For a molecule with $B_e/B_{\text{O}_2}>10$, the spin relaxation would be suppressed
by two orders of magnitude.
As the collision energy decreases, the variation of the cross sections with $B_e$ becomes more complicated.
At the lowest collision energy of $10^{-6}$ K, both the elastic and spin-changing cross sections decrease with increasing $B_e$,
and the suppression of spin relaxation in very efficient for large rotational constants. In contrast, for the intermediate kinetic energy
of 1 mK, the probability for spin relaxation remains large even for $B_e/B_{\text{O}_2}>10$, and the cross sections
show oscillations due to Feshbach resonances (see Fig. 4). The results shown in Figs. 7 and 8 indicate that off-resonant laser fields
may be used to suppress spin-changing transitions and induce Feshbach resonances in ultracold collisions of $^3\Sigma$ molecules.
We note that it order to establish that the laser-induced magnetic Feshbach resonances do occur at practicable laser field intensities,
it will be necessary
to perform accurate quantum scattering calculations in superimposed magnetic and off-resonant laser fields. Such calculations can be
carried out using the formalism presented in this work.

\section{Summary}

We have presented a quantum mechanical theory and converged quantum mechanical calculations for collisions of identical $^3\Sigma$ molecules
in a magnetic field. The cross sections for spin relaxation at magnetic fields above 1 mT are comparable to the elastic cross
sections, and they are not very sensitive to the applied field. At low magnetic fields, the spin-flipping transitions are suppressed
by $d$-wave centrifugal barriers in the outgoing collision channel, suggesting that $^3\Sigma$ molecules can be evaporatively
cooled in shallow magnetic traps. Our calculations shown in Fig. 4 suggest that the cross sections for collisions of molecules in the absolute ground state display a dense spectrum of Feshbach resonances, which may be used to tune molecule - molecule scattering lengths in a very wide range of magnetic fields.  
These resonances can be used to create tetra-atomic molecules via magnetoassociation \cite{Chin},
manipulate molecule-molecule interactions in optical lattices \cite{Bloch}, control chemical reactions of polyatomic molecules \cite{RomanReview},
or facilitate evaporative cooling of molecules in an optical dipole trap \cite{Cs}.
Our results show that the number of Feshbach resonances for molecule-molecule scattering in the absolute
ground state ($\sim100$ T$^{-1}$) is larger by a factor of 10 than the density of the resonances for the low-field-seeking states.

\begin{acknowledgments}
We acknowledge useful discussions with Kirk Madison, Nathan Brahms, and Simonetta Cavalli.
This work was supported by NSERC of Canada and an NSF grant to the Institute for Theoretical Atomic, Molecular, and Optical Physics
at Harvard University and the Smithsonian Astrophysical Observatory.
\end{acknowledgments} 

\appendix
\section{Matrix elements of the interaction potential}

 In principle, the radial coefficients $V_{\lA\lB\lambda}(R)$ can be
obtained by inverting Eq. (\ref{V}) using the orthonormality properties of
the spherical harmonics. This procedure requires the evaluation of six-dimensional
integrals over the angles ($\hR,\hrA,\hrB$), which poses a difficult practical problem. 
A more convenient method of evaluating the expansion coefficients in Eq. (\ref{V}) involves a transformation
to the body-fixed (BF) frame. The $z$-axis of
the BF frame coincides with the vector $\R$ \cite{Green,Kouri}. The tensor product in Eq. (\ref{A}) is 
invariant under rotations of the coordinate system \cite{Zare,Varshalovich}, so we
can substitute $\hR=\hat{0}$ into Eq. (\ref{V}) and use Eq. (\ref{A}) to obtain
\begin{align}\label{BF}\notag
V_\text{si}({\bm R},\hrA,\hrB)
= (4\pi)^{3/2}& \sum_{\lambda_\text{A},\lambda_\text{B},\lambda}
\left(\frac{2\lambda+1}{4\pi}\right)^{1/2} V_{\lA\lB\lambda}(R) \\ \times
&\sum_{m}\threejm{\lA}{m}{\lB}{-m}{\lambda}{0}Y_{\lA m}(\hrA')
Y_{\lB, -m}(\hrB'),
\end{align}
where the primes indicate the BF angles. In deriving Eq. (\ref{BF}) we have used the fact that
$Y_{\lambda m_\lambda}(\hat{0}) = [(2\lambda+1)/4\pi]^{1/2}$. We define the BF angular basis
functions \cite{Green,Kouri,JACS}
\begin{equation}\label{ABF}
A_{\lA\lB\lambda}(\hrA',\hrB')=(2\lambda+1)^{1/2}
\sum_{m}\threejm{\lA}{m}{\lB}{-m}{\lambda}{0}Y_{\lA m}(\hrA')
Y_{\lB, -m}(\hrB'),
\end{equation}
to expand the interaction potential as follows
\begin{equation}\label{VBF}
V_\text{si}({\bm R},\hrA,\hrB)
= 4\pi\sum_{\lambda_\text{A},\lambda_\text{B},\lambda}A_{\lA\lB\lambda}(\hrA',\hrB')V_{\lA\lB\lambda}(R).
\end{equation}
We note that the radial coefficients $V_{\lA\lB\lambda}(R)$ in Eqs. (\ref{VBF}) and (\ref{V}) are exactly the same.
This is not the case for the basis functions expressed in different coordinate systems. 

The BF basis functions span a coordinate subspace defined by 4 angles, $\hrA'=(\theta_\text{A},\varphi_\text{A}$)
and $\hrB'=(\theta_\text{B},\varphi_\text{B}$). However, the interaction potential depends on the dihedral
angle $\varphi = \varphi_\text{A}-\varphi_\text{B}$ and not on the azimuthal angles separately.
By separating the sum in Eq. (\ref{ABF}) into three contributions from $m>0$, $m=0$, and $m<0$,
respectively, and changing the sign of the summation indexes with $m<0$, we obtain
\begin{equation}\label{Real}
A_{\lA\lB\lambda}(\hrA',\hrB')=\frac{(2\lambda+1)^{1/2}}{2\pi}
\sum_{m=0}^{m_\text{max}} \threejm{\lA}{m}{\lB}{-m}{\lambda}{0}  
 \Theta_{\lA m} (\theta_\text{A})  \Theta_{\lB m} (\theta_\text{B}) [2-\delta_{m,0}](-)^m \cos (m\varphi),
\end{equation}
where $\Theta_{\lB, -m} (\theta_\text{B})$ are the normalized associated
Legendre polynomials. Here, we have used the relation $(-)^{\lA+\lB+\lambda}=+1$ that holds for two identical homonuclear molecules.

In practice, it is more convenient to use the basis functions that depend on the angles $\theta_\text{A}$, $\theta_\text{B}$, and $\varphi$.
Even though the right-hand side of Eq. (\ref{Real}) depends on the three angles, the functions on the left-hand side are defined in
a four-angle space. Therefore, we need to renormalize Eq. (\ref{Real}) to exclude the integration over one
extra angle. We define a reduced orthonormal basis set of functions, which span the three-dimensional subspace of angles
($\theta_\text{A},\theta_\text{B},\varphi$)
\begin{multline}\label{Real3}
\tilde{A}_{\lA\lB\lambda}(\theta_\text{A},\theta_\text{B},\varphi)=\frac{1}{\sqrt{2\pi}}A_{\lA\lB\lambda}(\hrA',\hrB') 
=\left(\frac{2\lambda+1}{2\pi}\right)^{1/2} \\ \times
\sum_{m=0}^{m_\text{max}} \threejm{\lA}{m}{\lB}{-m}{\lambda}{0}  
 \Theta_{\lA m} (\theta_\text{A})  \Theta_{\lB m} (\theta_\text{B}) [2-\delta_{m,0}](-)^m \cos (m\varphi)
\end{multline}
We emphasize that unlike Eq. (\ref{Real}), this expression defines an orthonormal basis in the
subspace of three independent angular variables. This suggests the following expansion
\begin{equation}\label{Vexp}
V_\text{si}(R,\theta_\text{A},\theta_\text{B},\varphi) = \sum_{\lA,\lB,\lambda}\tilde{V}_{\lA\lB \lambda}(R)\tilde{A}_{\lA\lB\lambda}(\theta_\text{A},\theta_\text{B},\varphi)
\end{equation}
We note that the coefficients $\tilde{V}_{\lA\lB \lambda}(R)$ are {\it not} the same as in Eq. (\ref{V}). 
Multiplying Eq. (\ref{Vexp}) by $\tilde{A}_{\lA'\lB'\lambda'}(\theta_\text{A},\theta_\text{B},\varphi)$
and integrating over all angles, we find
\begin{equation}\label{Vcoef}
\tilde{V}_{\lA\lB \lambda}(R) = \int_0^{2\pi}d\varphi \int_0^\pi d\theta_\text{A}\sin\theta_\text{A}
\int_0^\pi d\theta_\text{B}\sin\theta_\text{B} \tilde{A}_{\lA\lB\lambda}(\theta_\text{A},\theta_\text{B},\varphi)
V(R,\theta_\text{A},\theta_\text{B},\varphi)
\end{equation}
Comparing the expansion (\ref{Vexp}) with Eq. (\ref{VBF}) and using Eq. (\ref{Real3}), we obtain the relation
between the original coefficients defined by Eqs. (\ref{V}), (\ref{BF}), (\ref{VBF}) and those given by Eq. (\ref{Vcoef})
\begin{equation}\label{Vrelation}
V_{\lA\lB \lambda}(R) = \frac{1}{\sqrt{8\pi}} \tilde{V}_{\lA\lB \lambda}(R).
\end{equation}
To summarize, the expansion coefficients $V_{\lA\lB \lambda}(R)$ can be obtained by (i) evaluating
the reduced basis functions (\ref{Real3}); (ii) integrating the interaction potential with these functions following Eq. (\ref{Vcoef}),
and (iii) using Eq.~(\ref{Vrelation}) to obtain the original expansion coefficients  (\ref{V}).

\section{Differential scattering cross sections for identical molecules}

The differential cross section (DCS) for collisions of indistinguishable molecules can be expressed in terms of the symmetrized
scattering amplitude $\tilde{q}^\eta_{\gA\gB\to \gA'\gB'}(\hR_\i,\hR)$ given by Eq.~(\ref{qsymm})
\begin{equation}\label{DCS}\notag
\frac{d\sigma_{\gA\gB\to\gA'\gB'}}{d\hR_\i d\hR}(\hR_\i,\hR) = \frac{1}{k^2_{\gA\gB}}
| \tilde{q}^\eta_{\gA\gB\to \gA'\gB'}(\hR_\i,\hR) |^2.
\end{equation}
The integral cross section can be obtained by integrating the DCS over the coordinates of the final
flux and averaging over all possible directions of the initial flux. 
\begin{equation}\label{ICS}
\sigma_{\gA\gB\to\gA'\gB'} = \frac{1}{4\pi} \iint d\hR_\i d\hR \
\frac{d\sigma^\eta_{\gA\gB\to\gA'\gB'}}{d\hR_\i d\hR}(\hR_\i,\hR)
\end{equation}

Since the scattered molecules in the
same internal state are not distinguishable after the collision, the integration over $\hR$ in Eq. (\ref{ICS})
has to be restricted over half-space \cite{Dalibard,AvdeenkovBohn} for the final states satisfying $\gA'=\gB'$.
Thus, Eq. (\ref{ICS}) can be written as
\begin{equation}\label{ICSsymm}
\sigma_{\gA\gB\to\gA'\gB'} = \frac{1}{4\pi} \int_{0}^{2\pi} d\phi_\i \int_0^{\pi} \sin\theta_\i d\theta_\i 
\int_{0}^{2\pi} d\phi \int_0^{\theta_\text{max}} \sin\theta d\theta 
\frac{d\sigma^\eta_{\gA\gB\to\gA'\gB'}}{d\hR_\i d\hR}(\theta_\i,\phi_\i;\theta,\phi),
\end{equation}
where $\hR_\i = (\theta_\i,\phi_\i)$, $\hR = (\theta,\phi)$, and we have defined $\theta_\text{max}$
to be $\pi/2$ if $\gA'=\gB'$ and $\pi$ otherwise.  The integration in Eq. (\ref{ICSsymm})
can be performed trivially for $\theta_\text{max}=\pi$ to yield \cite{Kouri,Green}
\begin{equation}\label{ICStrivial}
\sigma_{\gA\gB\to\gA'\gB'} =
\frac{\pi (1+\delta_{\gA\gB})(1+\delta_{\gA'\gB'})}
{k^2_{\gA\gB}} \sum_{\ell,m_\ell}\sum_{\ell' m_\ell'}|T^\eta_{\gA\gB\ell m_\ell ; \gA'\gB'\ell' m_\ell'}|^2,
\, \text{if $\gA'\ne \gB'$}.
\end{equation}
All that remains is to consider the special case $\gA' = \gB'$. Expanding the square modulus
in Eq. (\ref{ICSsymm}) yields
\begin{multline}\label{ExpandedModulus}
\sigma_{\gA\gB\to\gA'\gB'} =
\frac{\pi (1+\delta_{\gA\gB})2}{k^2_{\gA\gB}}
\biggl{[}
\sum_{\ell_1,m_{\ell_1}}\sum_{\ell_2,m_{\ell_2}}
\sum_{\ell'_1,m'_{\ell_1}}\sum_{\ell'_2,m'_{\ell_2}} \i^{-\ell_1 + \ell_1' +\ell_2-\ell_2'} \\ \times
T^{\eta*}_{\gA\gB\ell_1 m_{\ell_1} ; \gA'\gB'\ell'_1 m'_{\ell_1}}
T^{\eta}_{\gA\gB\ell_2 m_{\ell_2} ; \gA'\gB'\ell'_2 m'_{\ell_2}} 
\int_{0}^{2\pi} d\phi_\i \int_0^{\pi} \sin\theta_\i d\theta_\i Y_{\ell_1 m_{\ell_1}}(\theta_\i,\phi_\i)
Y^*_{\ell_2 m_{\ell_2}}(\theta_\i,\phi_\i) \\ \times
\int_{0}^{2\pi} d\phi \int_0^{\pi/2} \sin\theta d\theta 
Y_{\ell_1' m'_{\ell_1}}(\theta,\phi)
Y^*_{\ell_2' m'_{\ell_2}}(\theta,\phi)
\biggr{]}.
\end{multline}
The first integral is the usual normalization integral of two spherical harmonics.
The second integral can be separated in two parts \cite{Zare}
\begin{align}\label{Integral1}\notag
\int_{0}^{2\pi} d\phi \int_0^{\pi/2} \sin\theta d\theta Y_{\ell_1' m'_{\ell_1}}(\theta,\phi)
Y^*_{\ell_2' m'_{\ell_2}}(\theta,\phi) &= 
\int_{0}^{2\pi} d\phi\Phi_{m'_{\ell_1}}(\phi) \Phi^*_{m'_{\ell_2}}(\phi) \\ &\times
\int_0^{\pi/2}\sin\theta d\theta \Theta_{\ell_1' m'_{\ell_1}}(\theta)\Theta_{\ell_2' m'_{\ell_2}}(\theta).
\end{align}
The one-dimensional integral over $\phi$ is equal to $\delta_{m'_{\ell_1}m'_{\ell_2}}$, so that
the intergal (\ref{Integral1}) becomes
\begin{equation}\label{Integral2}
\delta_{m'_{\ell_1}m'_{\ell_2}}
\int_0^{\pi/2}\sin\theta d\theta \Theta_{\ell_1' m'_{\ell_1}}(\theta)\Theta_{\ell_2' m'_{\ell_1}}(\theta).
\end{equation}
Since for identical molecules $\ell_1'$ and $\ell_2'$ are of the same parity (either
even or odd) \cite{footnote3}, and the index $m'_{\ell_1}$ is fixed, the associated Legendre polynomials 
in Eq. (\ref{Integral2}) are both even or odd functions of $\cos\theta$ \cite{Zare,Varshalovich}. Therefore, their product is always
an even function of $\cos\theta$. Since in this case the substitution $\theta\to \pi-\theta$ leaves the
integrand in Eq. (\ref{Integral2}) unchanged, we can halve the $\theta$ integration range and obtain
\begin{equation}\label{Integral3}
\int_{0}^{2\pi} d\phi \int_0^{\pi/2} \sin\theta d\theta Y_{\ell_1' m'_{\ell_1}}(\theta,\phi)
Y^*_{\ell_2' m'_{\ell_2}}(\theta,\phi)
=\textstyle{\frac{1}{2}}\delta_{\ell'_1\ell'_2} \delta_{m'_{\ell_1}m'_{\ell_2}}.
\end{equation}
Substituting this result into Eq. (\ref{ExpandedModulus}) and performing the integration
leads to the cancellation of the prefactor $(1+\delta_{\gA'\gB'})$ by the $\frac{1}{2}$ from
Eq. (\ref{Integral3}). This completes the derivation of Eq.~(\ref{ICSfinal}).

\begin{figure}
	\centering
	\includegraphics[width=0.6\textwidth, trim = 0 10 0 0]{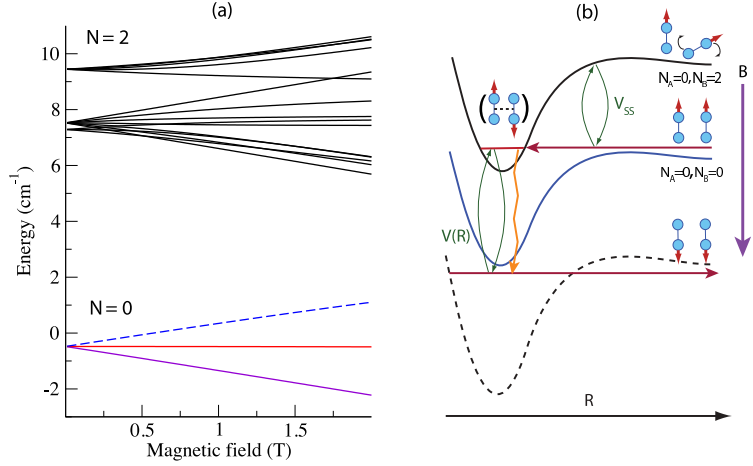}
	\renewcommand{\figurename}{Figure}
	\caption{($a$) Energy levels of the $^{17}$O$_2(^3\Sigma_g^-)$ molecule as functions of the magnetic field. ($b$) A schematic illustration
	of spin relaxation in collisions of $^3\Sigma$ molecules.}
\end{figure}

\begin{figure}
	\centering
	\includegraphics[width=0.45\textwidth, trim = 0 10 0 0]{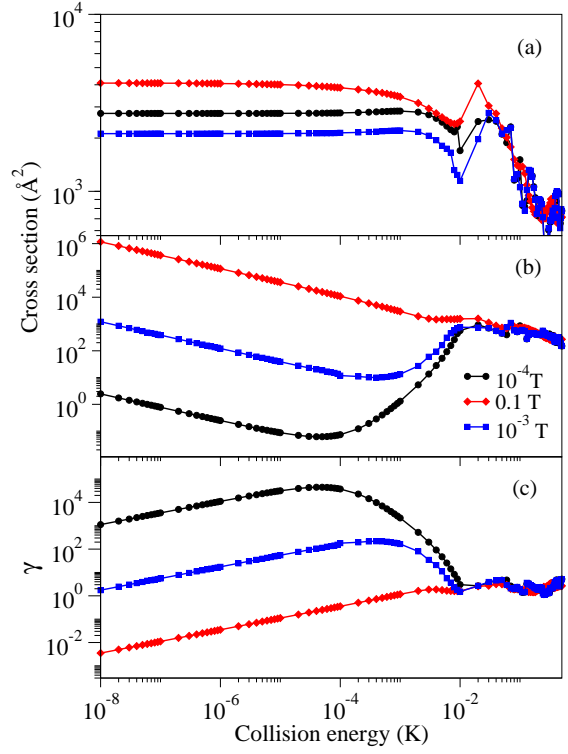}
	\renewcommand{\figurename}{Figure}
	\caption{The cross sections for elastic scattering ($a$) and spin relaxation ($b$) in O$_2(^3\Sigma_g^-)$ - O$_2(^3\Sigma_g^-)$ collisions
	as functions of the collision energy at different magnetic fields: $10^{-4}$ T (circles), 10$^{-3}$ T (squares),
	and 0.1 T (diamonds). The ratio of the cross sections for elastic scattering and spin
	relaxation is shown in panel ($c$).}
\end{figure}

\begin{figure}
	\centering
	\includegraphics[width=0.55\textwidth, trim = 0 0 0 0]{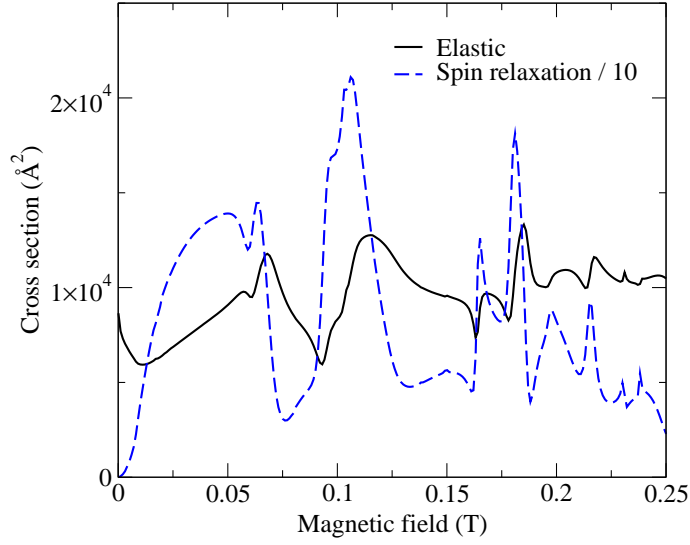}
	\renewcommand{\figurename}{Figure}
	\caption{The cross sections for elastic scattering (full line) and spin relaxation (dashed line) in collisions of O$_2(^3\Sigma_g^-)$ molecules
	in the low field-seeking state $|M_{S_\text{A}}=1,M_{S_\text{B}}=1\rangle$ as functions of the magnetic field.
	The spin relaxation cross section is summed over all final spin states and
       divided by 10 to fit the scale of the figure. The collision energy is $10^{-6}$ K. }
\end{figure}

\begin{figure}
	\centering
	\includegraphics[width=0.55\textwidth, trim = 0 0 0 0]{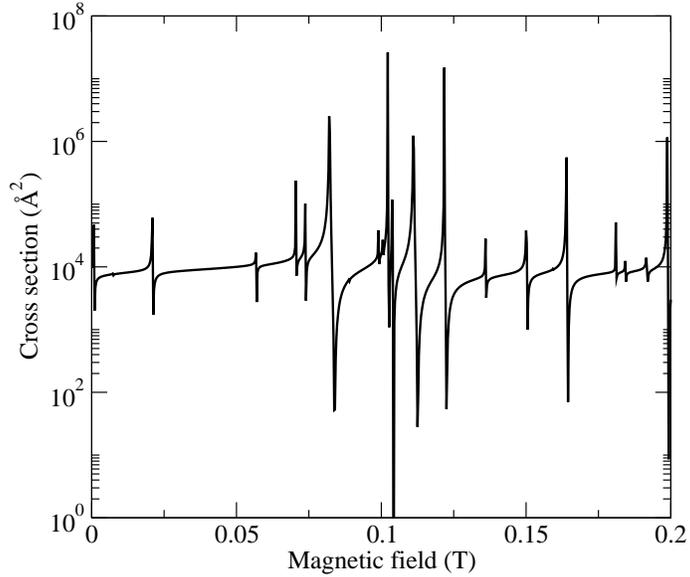}
	\renewcommand{\figurename}{Figure}
	\caption{The $s$-wave elastic scattering cross section for collisions	of O$_2(^3\Sigma_g^-)$ molecules in the lowest high-field-seeking state
	$|M_{S_\text{A}}=-1,M_{S_\text{B}}=-1\rangle$ as a function of the magnetic field. The collision energy is $10^{-6}$ K.}
\end{figure}

\begin{figure}
	\centering
	\includegraphics[width=0.6\textwidth, trim = 0 0 0 0]{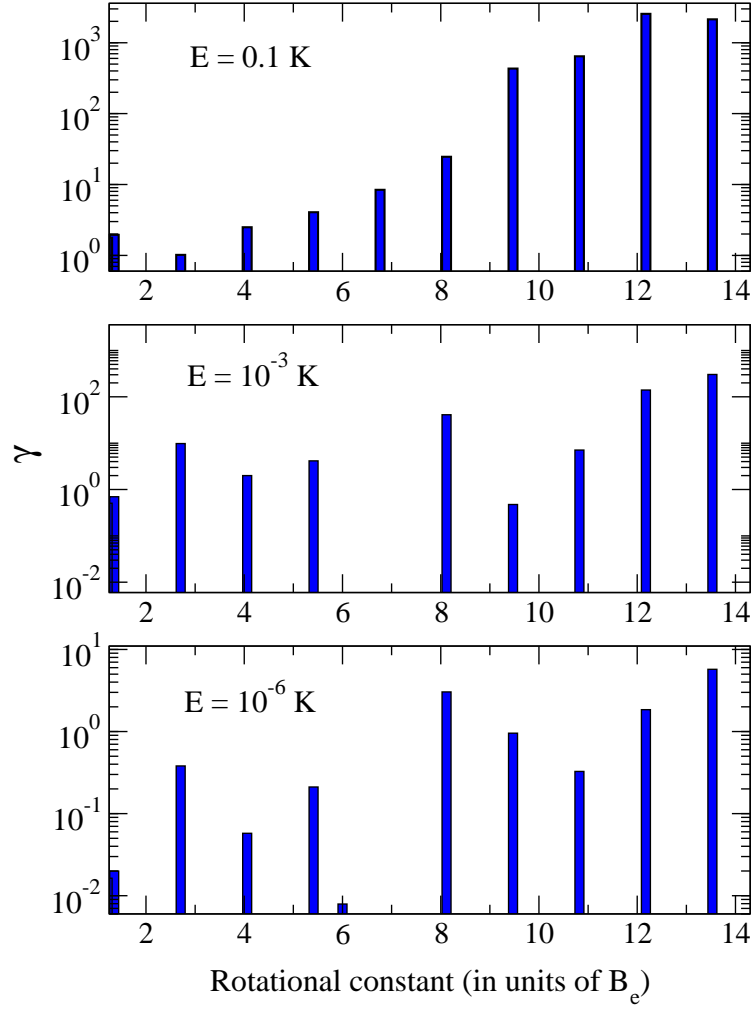}
	\renewcommand{\figurename}{Figure}
	\caption{The ratio of the cross sections for elastic scattering and spin relaxation in collisions of O$_2(^3\Sigma_g^-)$ molecules
	in the low-field-seeking state $|M_{S_\text{A}}=1,M_{S_\text{B}}=1\rangle$ as a function of the rotational constant at different
	 collision energies. The magnetic field is 1 T. }
\end{figure}

\begin{figure}
	\centering
	\includegraphics[width=0.6\textwidth, trim = 0 0 0 0]{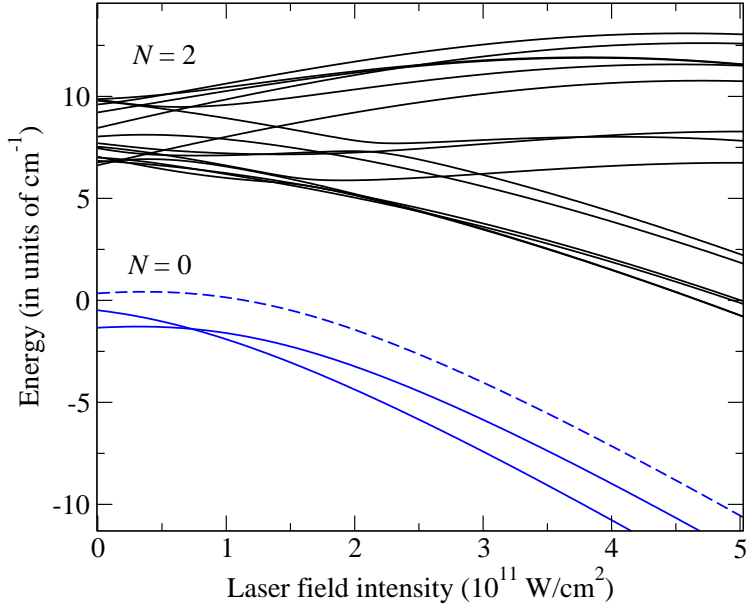}
	\renewcommand{\figurename}{Figure}
	\caption{Energy levels of $^{17}$O$_2(^3\Sigma^-_g)$ in superimposed magnetic and off-resonant laser fields.
	The levels are shown as functions of the laser field intensity at a fixed magnetic field of 1 T. The dashed line shows the
	highest low-field-seeking state $|M_{S_\text{A}}\rangle=1$. The laser field intensity is defined as $I_0 = \epsilon_0^2$ (see text for details).}
\end{figure}

\begin{figure}
	\centering
	\includegraphics[width=0.6\textwidth, trim = 0 0 0 0]{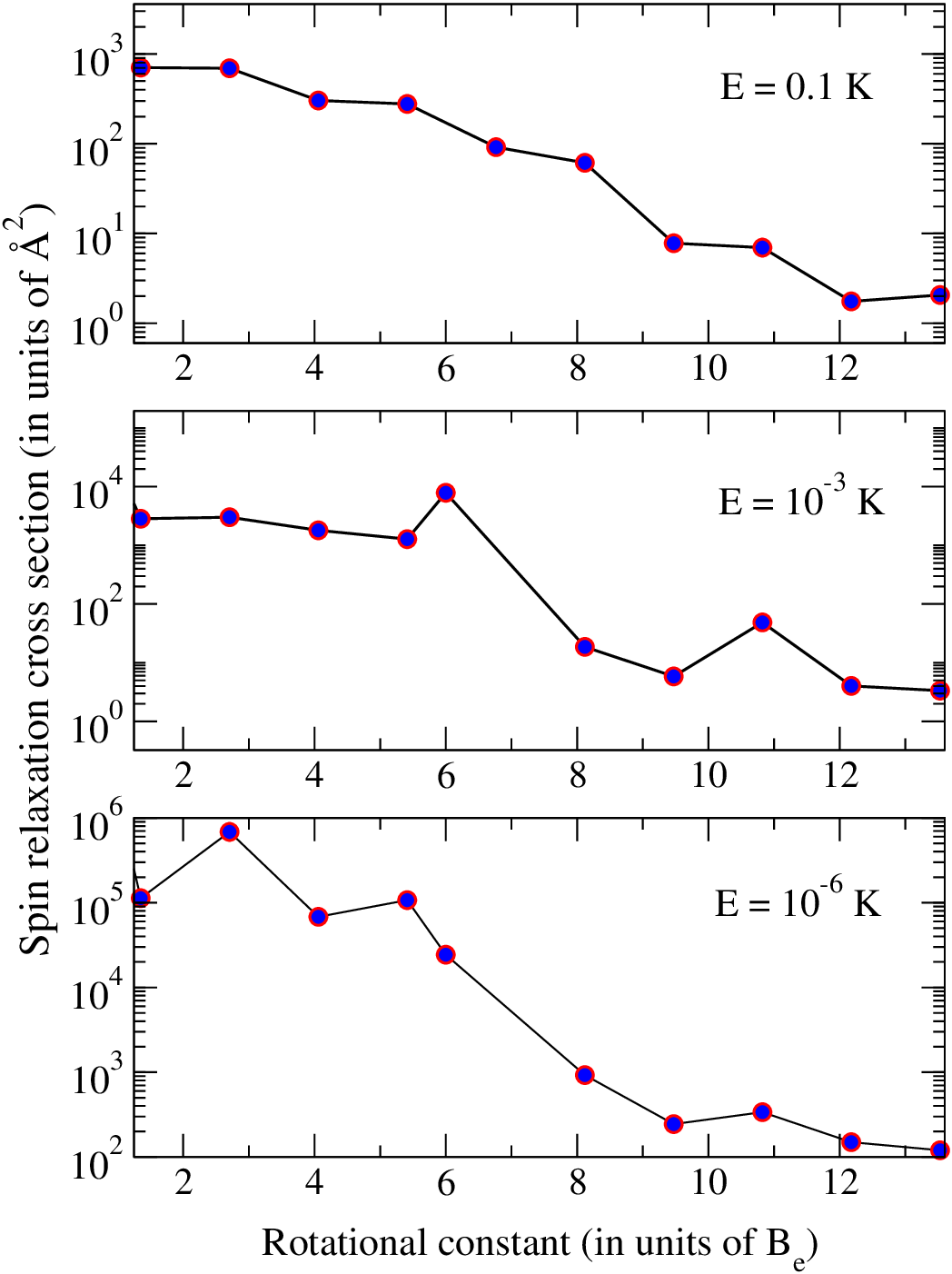}
	\renewcommand{\figurename}{Figure}
	\caption{The cross sections for spin relaxation in collisions of O$_2(^3\Sigma_g^-)$ molecules
	in the low-field-seeking state $|M_{S_\text{A}}=1,M_{S_\text{B}}=1\rangle$ as functions of the rotational constant at different
	 collision energies. The magnetic field is 1 T. }
\end{figure}

\begin{figure}
	\centering
	\includegraphics[width=0.6\textwidth, trim = 0 0 0 0]{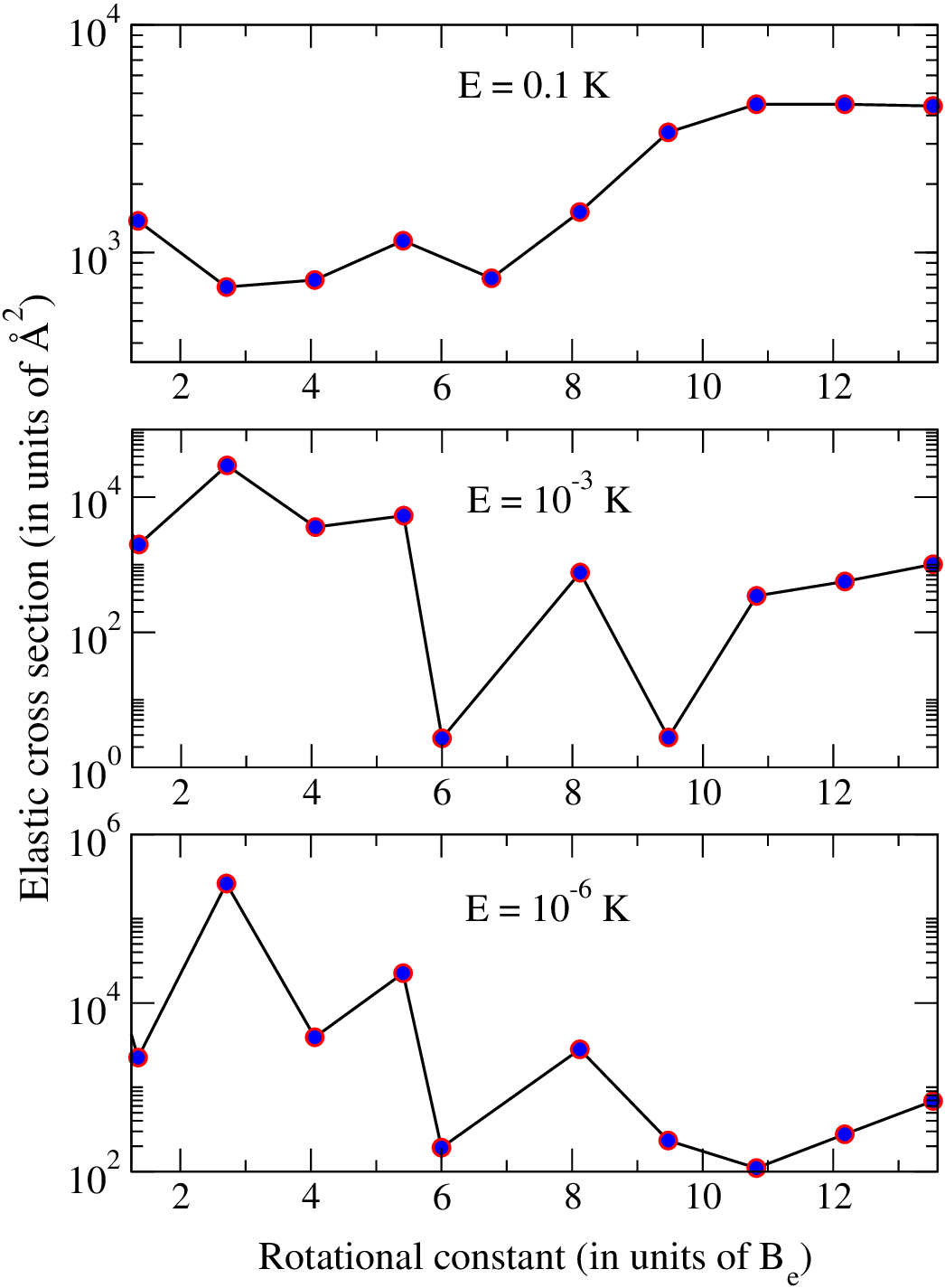}
	\renewcommand{\figurename}{Figure}
	\caption{The cross sections for elastic scattering in collisions of O$_2(^3\Sigma_g^-)$ molecules
	in the low-field-seeking state $|M_{S_\text{A}}=1,M_{S_\text{B}}=1\rangle$ as functions of the rotational constant at different
	 collision energies. The magnetic field is 1 T. }
\end{figure}


\begin{thebibliography}{99}
\bibitem{footnote1}
The index $S_\nu$ is omitted because the electron spin of the molecule is conserved in a collision.
\bibitem{footnote2}
The expansion of $J({\bm R},\rA,\rB)$ is identical to that of the interaction potential.
We therefore do not repeat it here.
\bibitem{footnote3} This applies to homonuclear as well as heteronuclear molecules. In both cases,
the states with $\gA=\gB$ are present only when $\eta(-)^{\ell}=+1$.
\end{thebibliography}
\end{document}